\documentclass[bibyear]{mfa}  

\usepackage{graphicx} 
\usepackage{url}
\usepackage{xspace}
\usepackage[T1]{fontenc}
\usepackage[utf8]{inputenc}
\usepackage{lmodern}
\usepackage[english]{babel}
\usepackage[autostyle]{csquotes}
\usepackage[toc,page]{appendix}
\usepackage{natbib,twoopt}
\usepackage{xcolor}
\usepackage[breaklinks=true]{hyperref} 

\begin{document} 

   \title{Conducting the SONG}
   
   \subtitle{The robotic nature and efficiency of a fully automated telescope}

   \author{M. Fredslund Andersen \inst{1}
   		\and
   			R. Handberg \inst{1}    
   		\and
   			E. Weiss \inst{1} 		
   		\and
   			S. Frandsen \inst{1} 	
   		\and
   			S. Sim\'on-D\'iaz \inst{2,3}    
   		\and
   			F. Grundahl \inst{1}  
   		\and
   			P. Pallé \inst{2,3}      }	

   \institute{Stellar Astrophysics Centre, Department of Physics and Astronomy, Aarhus University, 
Ny Munkegade 120, DK-8000 Aarhus C, Denmark
	\and
	Instituto de Astrofísica de Canarias. E-38205 La Laguna, Tenerife, Spain
	\and
	Departamento de Astrof\'isica, Universidad de La Laguna, E-38205 La Laguna, Tenerife, Spain}

   \date{Submitted: \today}

  \abstract 
   {We present a description of ``the Conductor'', an automated software package which handles all observations at the first Stellar Observations Network Group (SONG) node telescope at the Teide Observatory on the island of Tenerife. The idea was to provide a complete description on the automated procedures for target selection and execution of observations and to demonstrate how the SONG robotic telescope is being operated. The Conductor is a software package developed in Python and running on a server in Aarhus which makes use of a large set of database tables through which it communicates with the SONG nodes. Based on a number of selection criteria the Conductor is able to identify the optimum target to be observed at any given moment taking into account local weather conditions and technical constraints. The Conductor has made it possible for the Hertzsprung SONG telescope to become a highly sophisticated and efficient robotic telescopic facility without human interaction. It can handle everything from Principal Investigators submitting their proposed targets with specific settings to the data being available for download after the observations has been carried out. 
At present, and thanks to the availability of the Conductor, the first node of the SONG project can be considered a world leading robotic telescope with respect to needed human interactions, efficiency and flexibility in observing strategy.}


   \maketitle
%

\section{Introduction}
Within the last decade the technological development has made it possible to build and run  automated astronomical telescopes (e.g. \cite{stella},~\cite{lcogt},~\cite{liverpool}). The accessibility of network and computer controlled hardware components has increased enormously and most of the hardware needed for an automated telescope are of the shelf products.
The Stellar Observations Network Group (SONG) has been conceived as a global network of small 1\,m telescopes (\cite{old_song},~\cite{mfa1}). The network currently consists of the prototype (fully operational over the last 4 years) located at the Teide Observatory on the island of Tenerife. It was initiated through a collaboration between Aarhus University (AU), Copenhagen University (KU) and Instituto de Astrofísica de Canarias (IAC). A second node is currently being build at the Delingha observatory in China by the National Astronomical Observatories of China (NAOC). Close collaboration with the University of Southern Queensland (USQ), the University of Sydney (US), the University of New South Wales (UNSW), the Monash University (MU) and the Australian National University (ANU) has initiated the development of a third node at the Mount Kent Observatory in Australia and the building of the new fibre-fed spectrograph that will be attached to minimum two 70cm telescopes has recently started.\\
The Hertzsprung SONG telescope is equipped with a high resolution echelle spectrograph which uses an iodine cell for calibration when precise radial velocities are requested and has the possibility to use Thorium-Argon as well \citep{muher}.

\section{The essentials - The Teide Observatory}

When building a robotic observatory a few essentials are needed on site. Key requirements for site selection are a well-functioning infrastructure with a reliable power grid and a stable and fast Internet connection. Another critical component is on-site expertise to assist in case of sudden urgent problems and for preventive maintenance work. These key ingredients together with one of the best locations for astronomical observations has proven the Teide Observatory to be an optimum location to set up an autonomous and robotic observatory. The prototype were from a very early stage designed to become a robotic facility which meant all hardware components was selected with remote and automatic control in mind.

\begin{figure}[t!]
	\centering
    \includegraphics[width=\columnwidth]{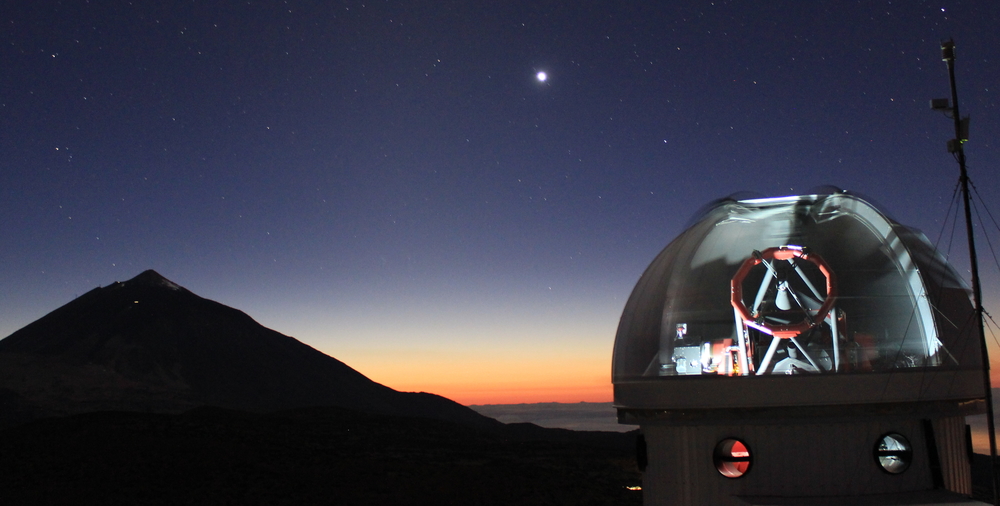}
    \caption{The Hertzsprung SONG telescope at the Teide Observatory in Tenerife. In the background the iconic volcano Teide is seen.}
    \label{fig:SONG_site}
\end{figure}

\section{The foundation - SODA}
\label{sec:SODA}
All SONG telescopes are managed and operated from Aarhus University, Denmark (Central), where a number of servers are set up to mainly run the SONG Data Archive (SODA)\footnote{\url{https://soda.phys.au.dk}} which acts as the foundation on which the automatic nature of the SONG observations is build. SODA has been developed to be a multi-purpose database system which serves as the access point for the users to:
\begin{itemize}
\item submit SONG observing proposals.
\item enter targets from observing proposals awarded observing time by the SONG Time Allocation Committee (TAC).
\item download data collected by any SONG telescope.\\
\end{itemize}

The SONG network has application calls twice a year where collaborators and project members are encouraged to submit proposals (Phase 1). The PI of a successful application will get access to Phase 2 in the application process. In this phase, the user will need to enter the targets to be observed with all relevant settings and constraints. The entering is done through a dedicated web form on the SODA web page. After submission in Phase 2 everything can run fully automatically until the data are accessible in the data archive.\\
SODA is built on the same layout as the KASOC and TASOC databases \citep{kasoc1}, \citep{kasoc2} which have proven to be stable and well-functioning systems. These databases house all data from the Kepler and the TESS missions and thereby gives the opportunity to do cross-talk between the databases to check for available data observed with the different facilities. The most critical database table in SODA for the automation of the SONG observations are shown in Appendix~\ref{A1}.

\section{The brain - The Conductor}
The night scheduling for the SONG node in Tenerife can be done in two ways. One is entering the targets by hand through a web interface. This web interface will only include targets from projects which have been allocated time in the given period. Targets entered here will have highest priority during the night. This interface is currently mainly used for technical and test observations. The other way for target selection is carried out by the \textbf{``Conductor''}. This Python based software package will run through all targets in SODA which have been marked as active in the given period and based on the settings defined by the PIs it will select the best target to observe at any given time during the night.

Selecting the right target to be observed at any given time is not a trivial task \citep{lcogt2}, \citep{stella2}. Each target inserted into SODA through Phase 2 will be defined by a number of parameters. The targets will at first be divided into different scheduling types:
\begin{enumerate}
  \item \textbf{Time Critical.} This group includes transits, eclipses or other targets to be observed at a specific date and time.
  \item \textbf{Radial Velocity Standard.} Those stars to be observed regularly to characterize and check the stability of the spectrograph. 
  \item \textbf{Large Program.} This type will be selected for targets where high cadence observations are needed during multiple nights.
  \item \textbf{Periodical.} The periodical targets are those that require one or a few exposures regularly with a roughly fixed time interval in between visits which could be nightly or weekly. An example of this kind could be exoplanet follow up with radial velocity measurements.
  \item \textbf{Filler.} Single or few exposures of single targets which can be acquired at any time.
  \item \textbf{Backup.} Targets that only are observed if no target was selected from the previous types. The backup targets are long term monitoring programs like binary stars.
\end{enumerate}

Depending on the scheduling type selected for a specific target different parameters needs to be set. In the case of Time Critical targets the PI must specify the start and stop time of the observations and all the relevant instrumental settings. The Large Program and Periodical targets will be defined by a ``time between observations'' ($\Delta t_\mathrm{base}$) and the rest has no specific parameters other than the standard instrumental settings, number of exposures per visit and minimum altitude of the observation (airmass).

\begin{figure}
	\centering
    \includegraphics[width=\columnwidth]{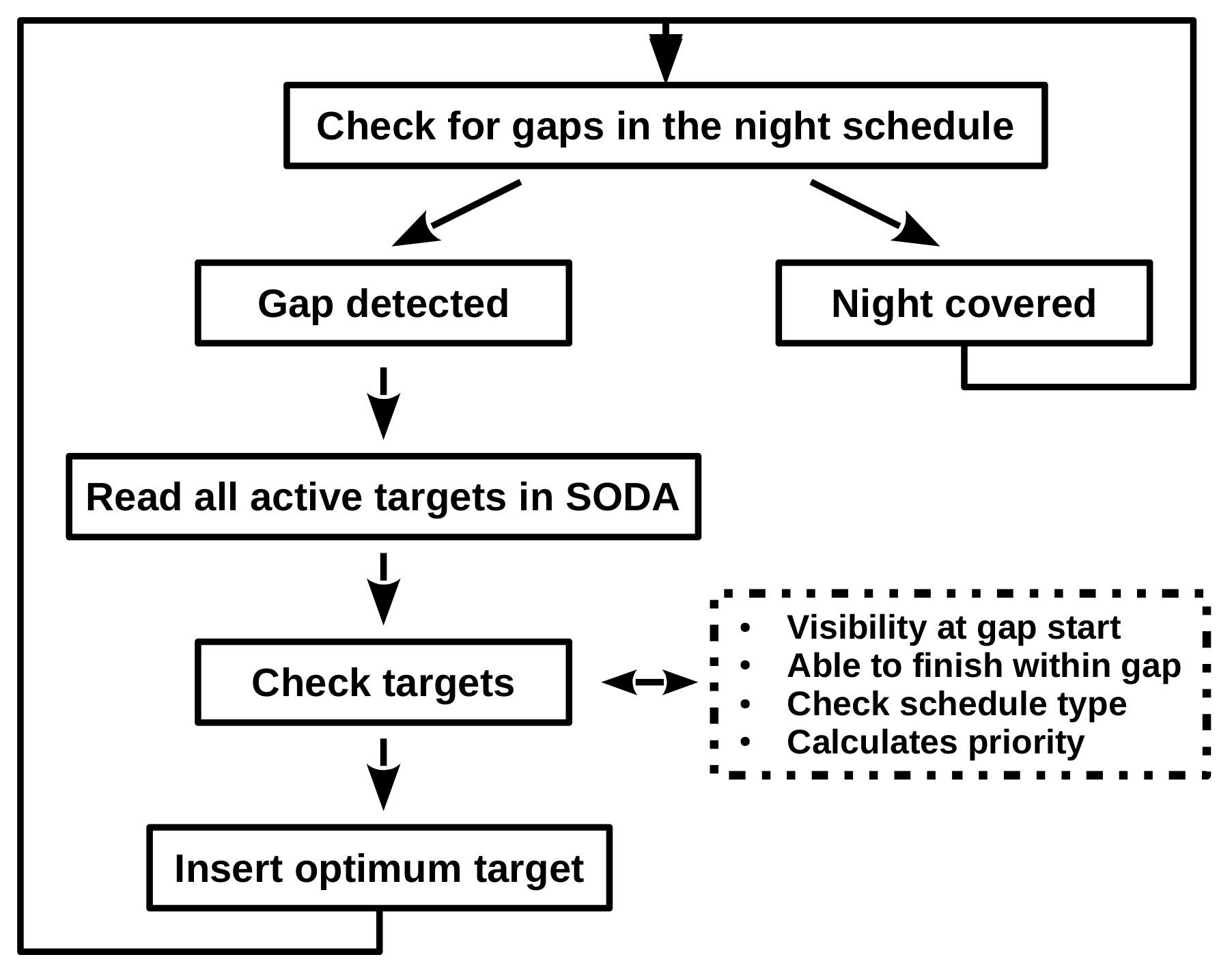}
    \caption{Flow diagram of the target selection procedure in the Conductor.}
    \label{fig:conductor_diagram}
\end{figure}

The Conductor will go through each scheduling type in a prioritized order (provided in the list above). Time Critical observations will be inserted as Observing Requests (OR) by the Conductor in the morning the day they should be executed and SONG staff members monitoring the facility daily will be notified by e-mail about the insertion. The other scheduling types are handled in real time. See Fig.~\ref{fig:conductor_diagram} for a flow diagram of the idea behind the selection procedure of the Conductor. 

The procedure followed by the Conductor is to check the schedule for the ongoing night and look for gaps in the observing program. If a detected gap is the entire night and the Sun is getting sufficiently low (6 degrees below horizon) the Conductor will insert the OR of the top prioritized target which it has identified to be observed from the beginning of the night.\\ 
The Conductor will run through the different scheduling types starting with the Radial Velocity Standard stars. Each of these which are observable at the given time will be observed roughly  once per night. Next it will check for Large Programs. In this category an ongoing asteroseismic campaign on a specific target would be defined and only one project at a time should be set to active. If no Large Program is selected it will go to the Periodical scheduling type. This type usually contains a large number of active targets and a priority will be calculated for each target according to a number of parameters. First targets are checked for visibility in the window defined by the exposure time (defined by the user) and only targets which can finish an exposure during the detected program gap will be selected. Other parameters like Moon distance and wind speed in the direction of the object will also be checked for each target. All targets that fulfill these requirements will be given a priority which will be calculated in the following way:
  \begin{equation}\label{eqn:periodical}
      p_\mathrm{periodical} = \frac{\Delta t}{\Delta t_\mathrm{base}} \cdot 100 \,
  \end{equation}
where $\Delta t_\mathrm{base}$ is the user defined requested time between observations and $\Delta t$ is the actual time since last observation of this target. If the target has not been observed before $\Delta t$ will be calculated from the start date of the ongoing observing period. The priority will be a value between 0 and 100 until $\Delta t$ increases beyond the defined $\Delta t_\mathrm{base}$. The priority will continue to increase (beyond 100) until it is observed again. When all targets in this schedule type have been given a priority, the Conductor will check if any has a priority above 90 and if so it will select the one with the highest value to be executed. If no target has been selected it will continue to the filler targets. Here the priority is calculated in a different way. First it loops through all observable targets and derives a preliminary priority which is given by: 
\begin{equation}
	p_\mathrm{pre} = 90 + 10\cdot \frac{1}{r_\mathrm{project}}
    \label{Eq:pre}
\end{equation}

All fillers which are observable at the given time will get a priority of more than 90 which means they can be selected and executed by the Conductor. The $r_\mathrm{project}$ is the rank of the projects defined by the SONG TAC based on their science evaluation (1 for highest rank, 2 for second highest and so forth). Highly ranked projects will be given a higher priority. This is just a short listing which might end up in multiple targets given the same high priority and another priority will then be added based on the objects position in the sky. 
To optimize the data quality and success rate in target acquisition the priority is determined in the following way:
\begin{equation}
	p_\mathrm{filler} =  p_\mathrm{pre} + \frac{f}{|h - 50| + 1} \,
    \label{Eq:pre_fill}
\end{equation}
$h$ is the altitude of the object at the specific time and $f$ is either a factor of $10$ or $20$ depending if the target is rising or setting respectively. This calculation will favour objects in an area of the sky where the target acquisition will most likely succeed every time (Alt/Az mounts have problems near zenith). It will also favour targets which are setting. The principle behind this approach is that there might be time afterwards to execute more targets where rising targets will still be observable. As an example two stars from the same project could from Eq.~\ref{Eq:pre} have the same priority but if one star is rising and the other one is setting and both are at 50 degrees in altitude the one setting would be selected based on Eq.~\ref{Eq:pre_fill}. If both are setting and one is at 70 degrees in altitude and the other one is at 60 degrees the one lowest will be selected. On the other hand if both are setting and one is at 20 degrees in altitude and the other is at 40 degrees the one at 40 degrees will be selected.  

If no target has been selected so far, a Backup list of long term targets will be checked in the same way as for the targets in the Periodical type (Eq.~\ref{eqn:periodical}) and the target with the highest priority will be selected. The original idea behind the Backup program was to ensure observations even when weather conditions were not perfect. In case of strong wind observations downwind could still be allowed. If no targets from the time allocated programs are available downwind some backup targets could be observed.\\
The Conductor is intended to be adaptable at any telescope which has the fundamental database setup mentioned in Section~\ref{sec:SODA}. With the underlying structure in place the only changes to adopt to a different telescope and instrument should be made in the dedicated configuration file. The most relevant parameters specified in this file are listed in Appendix~\ref{A2} where a schematic diagram of the most relevant components of the software package is also shown (Fig.~\ref{fig:conductor_setup}).
The full software package of the Conductor is planned to be submitted to GitHub\footnote{https://github.com} in the near future and will be available to the whole community. 
 
\section{The workers - The Scheduler and The Monitor}
When a target is selected by the Conductor it will be inserted as an OR into a specific table in a database at Central (AU). This database table (shown in Appendix~\ref{A3}) will then be copied to a server at the node by the database replication software SLONY\footnote{Will be obsolete with planned upgrade to PostgreSQL 10} (see \cite{mfa1} for more details). On the local server on site a software package called the \textbf{``Scheduler''} then checks the list of scheduled targets to be observed the given night. If only one target is listed for a given time it will check visibility and general observing conditions and if all parameters are fine it will execute the corresponding observing script. This observing script will read all specified instrumental settings along with the parameters of the object and start the sequence of slewing the telescope to the object, make acquisition of the target to put it on the slit, move motors for the instrumental setup and then at the end to proceed to acquire the defined number of spectra. After completion it will change the status of the observation from being ``executing'' to be ``done'' and the script will finish. A new target should then already be inserted in the queue and the Scheduler will execute the next observation.\\
Alongside the Scheduler another on-site software program is running called the \textbf{``Monitor''}. This package will check all relevant site conditions; weather, server status, stability temperature sensors, the Sun's altitude, telescope and dome status and position etc. If changing conditions are detected it will act accordingly i.e. if weather turns bad it will close the dome and power off the telescope, in the morning when the Sun is about to rise it will also close down, in the evening it will power on the telescope and open up the dome to be ready for observing if conditions allow and so on. If issues occur which the Monitor cannot solve on its own it will ask for help by sending an e-mail to the SONG staff members monitoring the facility daily. In case the issue is of critical character the Monitor will also send a text message to inform and even wake up staff if necessary. 

\section{Nightly flexibility and efficiency of the Conductor}
\begin{figure}
	\centering
    \includegraphics[width=\columnwidth]{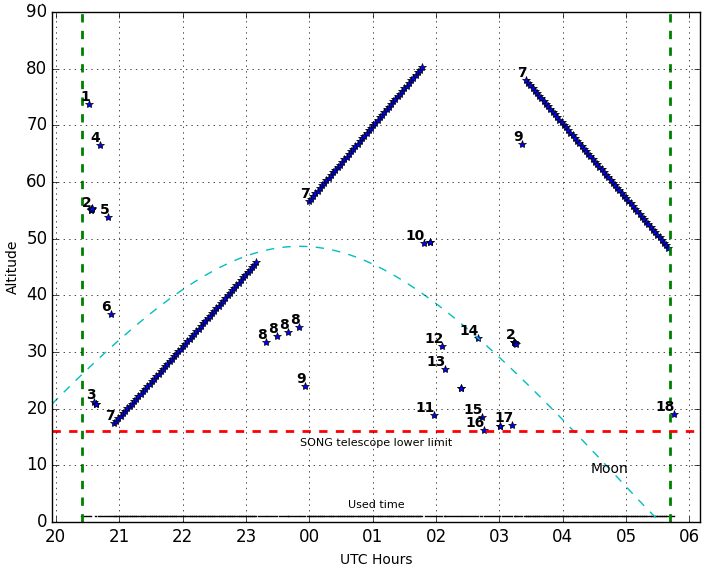}
    \caption{Observational status plot of the night 2018-05-27. Each asterisk marks an acquired spectrum for a given star which was located at the specific altitude at the specific time. Each number identifies individual stars. This night the main target (\#7) of scheduling type ``Large Program'' was using most of the telescope time. The majority of the other targets were ``Periodical'' with single exposures or ``Radial Velocity Standard'' stars. The vertical green dashed lines are marking the beginning and end time of the possible observing hours where the Sun is below 6 degrees under the horizon. The dashed cyan line is marking the path of the Moon during the given night. The dashed red horizontal line marks the altitude limit of the SONG telescope at Tenerife and the bottom black line labelled ``Used time'' is the actual exposure time of the CCD detector at the SONG spectrograph.}
    \label{fig:conducted_obs}
\end{figure}

In late spring 2017 the beta version of the Conductor was implemented and after a short period of testing and upgrading it was in a stage where it was handling most of the observations during summer 2017. Further development made it handle all observations from Autumn 2017 and onwards. 
An example of a full night where the Conductor selected all targets to observe is shown in Fig.~\ref{fig:conducted_obs}.

Each observed spectrum is marked on the figure with an asterisk at the specific time of acquisition where the object was at the plotted altitude in the sky. The numbering identifies different targets. In this particular night the object labelled number 7 was using most of the telescope time. This was a "Large Program" target observed for asteroseismology. The characteristics of this program allowed a few holes/gaps in the time series where other targets could be observed. Each block is defined by the number of exposures per visit and the requested time between visits. 

The total possible observing time this night was 09h:15m:00s, the CCD detector was exposed for 08h:26m:15s and the total readout time for the whole night was 7 minutes and 41 seconds (00h:07m:41s). The telescope was slewing 25 times with an overhead per target of 1 minute and 47 seconds (readout time excluded). During more than 90 \% of the total possible observing time the CCD detector was collecting science data. The overhead includes slewing the telescope, adjusting the instrumental setup and doing the target acquisition for the spectrograph.

\begin{figure}
	\centering
    \includegraphics[width=\columnwidth]{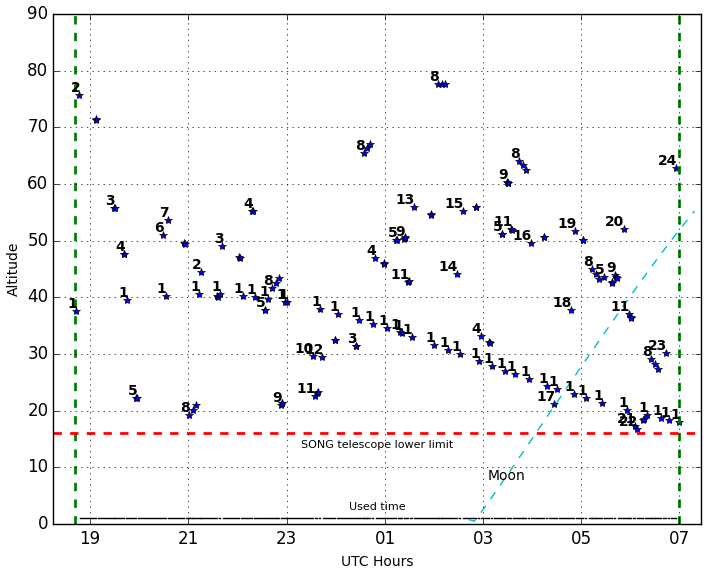}
    \caption{Another observational status plot of an entire night (2017-11-12) where the flexibility of the Conductor is really evident. This night a number of asteroseismic targets were observed with a user defined cadence (Periodical type). The required frequency of observations allowed multiple targets of the same kind and a number of filler exposures as well.}
    \label{fig:conductor_flexibility}
\end{figure}

Another example of the flexibility and efficiency of the Conductor and the SONG telescope is shown in Fig.~\ref{fig:conductor_flexibility}. This night was devoted to a number of asteroseismic targets (1, 3, 4, 5, 8, 9, 11) which could be observed at a cadence (Periodical) where switching between them were making full use of the observing time. Even though the telescope was slewing more than 80 times this night the time efficiency of exposing the CCD detector was still at an impressive 78 \% of the total possible observing window.  

\section{General efficiency and duty cycle}

\begin{figure*}
	\centering
    \includegraphics[width=\textwidth]{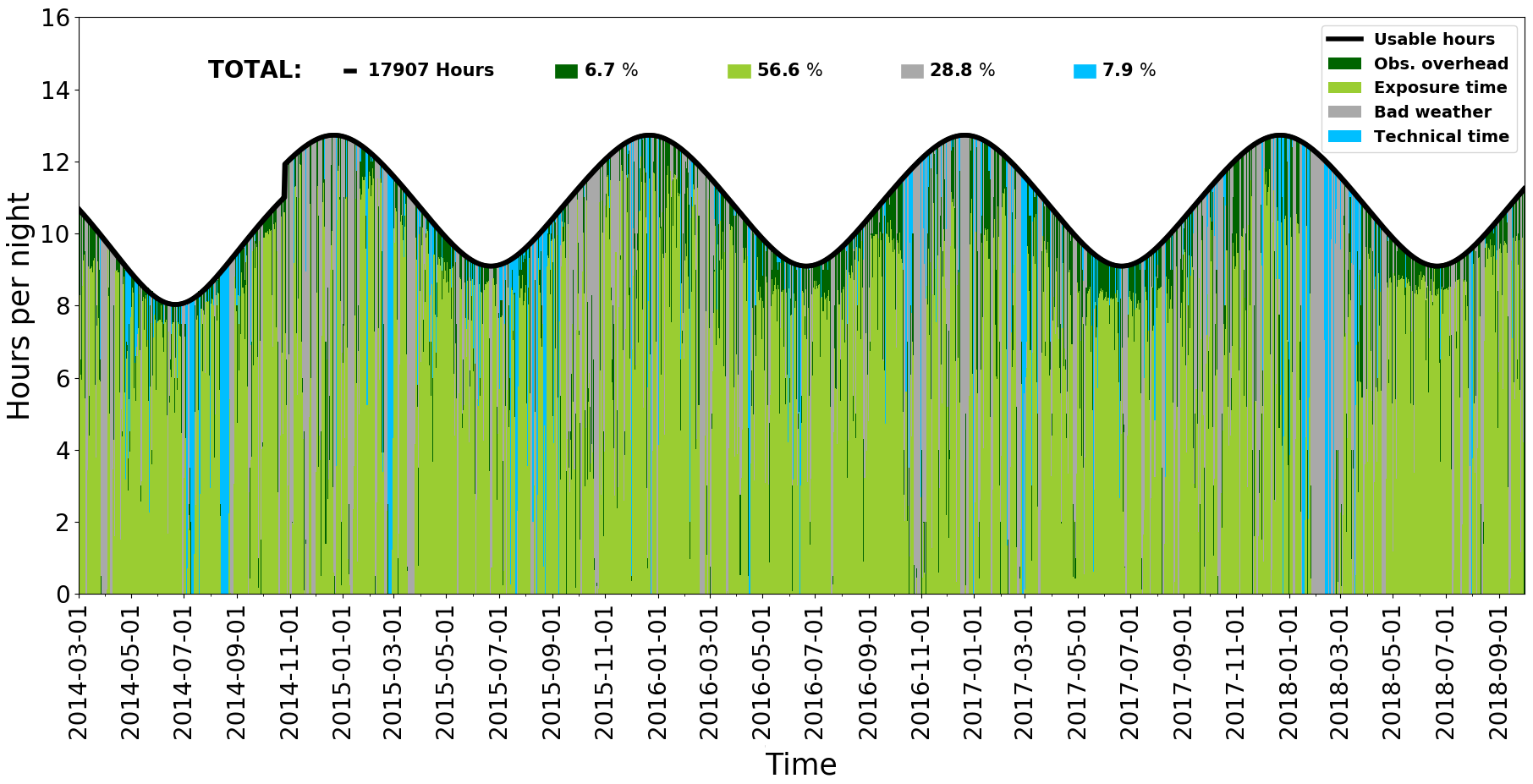}
    \caption{The total duty cycle of the first SONG node since scientific operations started in early 2014. The solid black line shows the possible observing hours each night. A clear change occurred around 2014-11-01 where the limits for observing was changed from end of nautical twilight (Sun altitude = -12$^{\circ}$) to end of civil twilight (Sun altitude = -6$^{\circ}$). The dark green color shows the observational overhead which includes slewing of the telescope, target acquisition, readout time etc. The light green color shows the part which was used for exposing the CCD. The gray color shows when the telescope remained closed because of bad weather. The blue color shows technical downtime. In the top part the total possible observing hours are shown and the percentage of this used on; observational overhead, exposing the CCD, bad weather and technical time is shown.}
    \label{fig:duty_cycle}
\end{figure*}

Minimizing the overhead and optimizing the observational procedures is one of the main goals of a fully robotic telescope. This leads to a highly efficient telescope on a nightly time scale. A downside on having an autonomous site is however that the telescope is often placed on a remote mountaintop far away from the operation center. This may result in long periods of downtime when a technical failure of some sort occurs and on-site work is needed.\\
In Fig.~\ref{fig:duty_cycle} the operational statistics since the beginning of the scientific operation in 2014 are shown. For each day five values are plotted; the length of the possible observing window in hours (black curve), the time used for exposing the CCD detector (light green), the observational overhead (dark green), downtime caused by bad weather (gray) and downtime due to technical work (blue). Technical time during bad weather was counted as ``bad weather'' since we often try to schedule preventive maintenance work and technical upgrades during bad weather. This also means that the value of about 29 \% of bad weather is an absolute value showing weather statistics of the site. During the time since 2014 some events are worth mentioning: 
In Autumn 2014 we decided to change the possible observing window of each night from starting at the end of nautical twilight to start at the end of civil twilight and also extending in the same way in the morning. This is seen as the increase in the total possible observing hours from November 2014. In August 2014 the dome rollers needed to be replaced causing 10 days of downtime. In February 2015 a major reconfiguration of the setup at the telescope nasmyth foci were performed. A major upgrade of the telescope control software and general maintenance of the dome and the telescope was performed by the telescope manufacturer (ASTELCO Systems GmbH) in late February 2017. An exceptionally extreme winter in 2018 (January \& February) caused a number of problems including a damaged weather station.\\
It is worth mentioning that 1.1 \% of the yearly technical downtime is around Christmas and new year where no night or daytime operators are present at the Teide Observatory and we decide to keep SONG closed independently of weather conditions or technical problems. If something fails we would not be able to ask someone on site for needed assistance to secure the hardware. Included in the technical time is also time spent on testing new instruments, creating pointing models, public outreach etc.\\
In the top of Fig.~\ref{fig:duty_cycle} the total number of possible observing hours are shown. Out of these hours 28.8 \% were not used caused by bad weather and 7.9 \% were devoted to technical work. 63.3 \% were used for observing including all overhead (slewing, target acquisition, readout, instrumental setup etc.) and the total time where the CCD detector was illuminated with star light was 56.6 \%. This means we on average have 6.7 \% ($\sim$ 1 hour) observational overhead each night which is to some extend visible in the figure during the long periods of good weather in spring and summer each year.

\section{The future - Multi-site conducting}
\begin{figure}[!ht]
	\centering
    \includegraphics[width=\columnwidth]{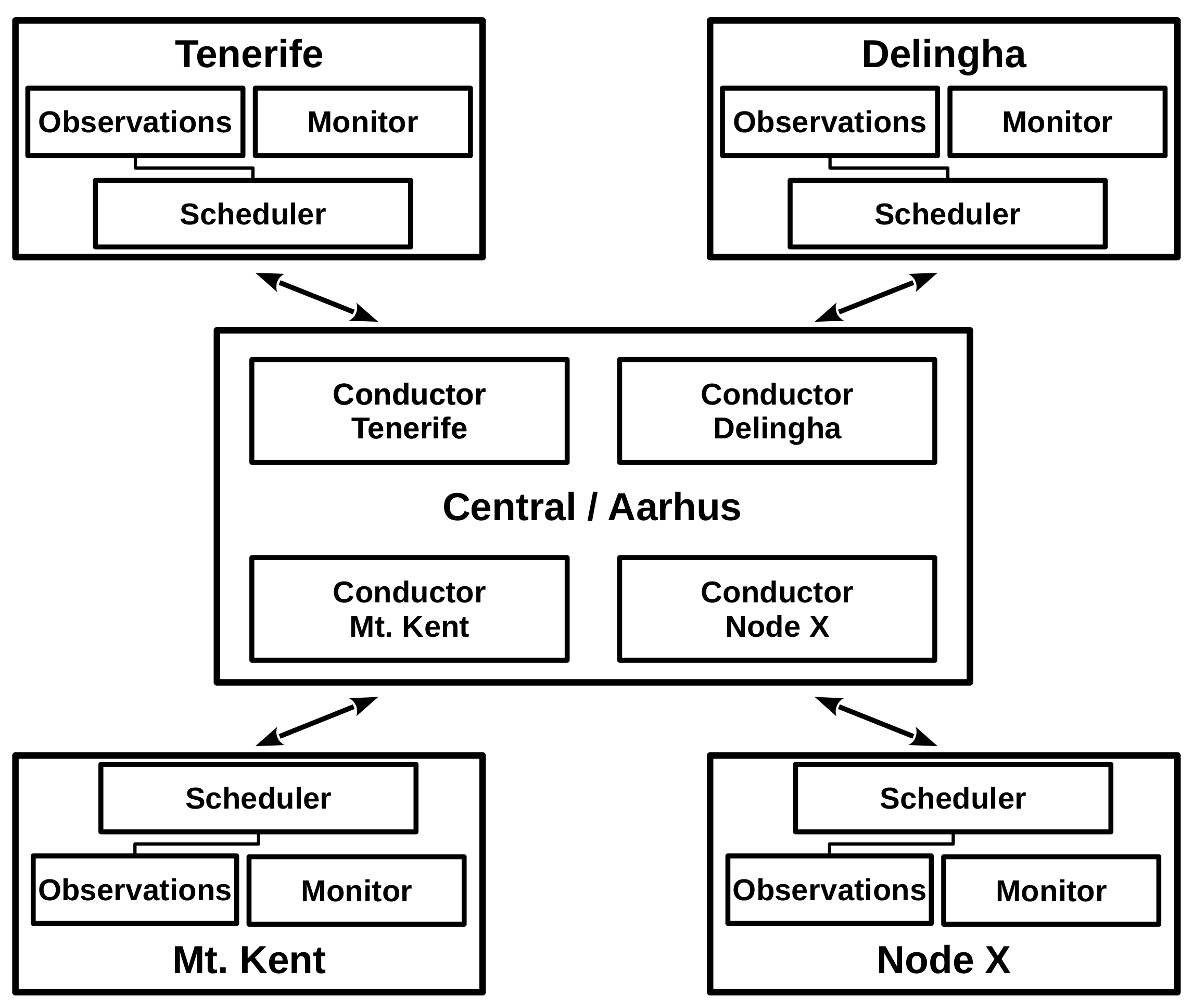}
    \caption{Schematic of multi-site conducting. The Conductors will run at Central in Aarhus and feed each node with targets to observe. At each node the Scheduler will execute the observations and report back to Central through the network replicated database tables.}
    \label{fig:multi-conductor}
\end{figure}

Operating the prototype SONG telescope at the Teide Observatory and having the Conductor running at Central has already proven to be a highly efficient and robust way of managing and executing the observations. Additional SONG nodes are being build and preparations to conduct the new sites in the same way as the prototype are well established. New nodes might have other constraints on the various observing and instrumental setup. Therefore, a Conductor for each SONG node will be implemented and all will run at Central. The general setup of multi-site conducting is visualized in Fig.~\ref{fig:multi-conductor}. At Central they will read and update the same database tables and will constantly have information from the sites on the status of every target being observed from the time allocated observing programs. 

\section{Further learning - Optimization}
The Conductor is running locally on a server at Aarhus University from where it reads the status from the Tenerife site through a network replicated database. The selected targets to be observed and inserted in the program queue is replicated to the site at Tenerife. Therefore the Conductor fully relies on a stable and fast network connection. In principle we do not need a fallback option since a local network failure should force a telescope shutdown. In case of a mechanical problem during the network failure, no remote inspection or actions can be carried out. There could however be situations where the network failure happens at the central location in Aarhus only causing problems for the Conductor and nothing else. In this case a fallback is relevant and a few possibilities are proposed:
\begin{itemize}
\item During a network failure the Scheduler could perform the same selection procedures as the Conductor but on local database tables instead. These tables would then be two-way replications of the relevant tables from the central database which will be up-to-date until the network failure. This would lead to a more optimized site which would still produce spectroscopic data during a network failure. A problem with this option is the lack of reply back to Central with the status of selected observations during the network outage. In this case, more telescopes (future nodes) might observe the same targets.
\item Another possibility is to have a fallback to a fixed set of backup targets. A simple and fixed selection procedure could then be implemented in an easy way. 
\item To carry on observing using an attached piggy back telescope for public outreach.
\end{itemize}

Another planned future software addition is to implement a fast response mechanism which will allow reaction time at a level of less than a minute from the trigger event to have the target on the spectrograph slit, and exposing the CCD detector. This addition will be named the ``Interrupter'' and will run locally as a software daemon listening for triggers.

\section{Conclusions}

We have described and demonstrated the efficiency and flexibility of the Hertzsprung SONG telescope which has become a fully robotic facility with the development of the software package ``the Conductor''. This software has turned the SONG site into a fully automated facility with minimum human interactions needed. Using simple priority calculation routines described here we show which parameters and how they are used when the Conductor is selecting objects to observe at any given time. The Conductor is not the only running software and the interplay with the Scheduler and the Monitor is essential for the site to operate as robust and efficient as we have now demonstrated.\\
The Conductor has been, from late 2017, handling all observations and has proven to be highly efficient, flexible and stable. Based on the results and experience during this period the Conductor will be adopted at all future SONG nodes.\\
Further development has been proposed to optimize the up-time even more and the incorporation of additional nodes into the network and software procedures are set for the near future. 

\begin{acknowledgements}
Based on data obtained at the Hertzsprung SONG telescope operated on the Spanish Observatorio del Teide on the island of Tenerife by the Aarhus and Copenhagen Universities and by the Instituto de Astrofísica de Canarias. Support for the construction and maintenance of the Hertzsprung SONG telescope from the Instituto de Astrofísica de Canarias, the Villum Foundation, the Carlsberg Foundation, the Independent Research Fund Denmark and the operator and maintenance teams of the Observatorio del
Teide is gratefully acknowledged. Funding for the Stellar Astrophysics Centre is provided by The Danish National Research Foundation (Grant agreement no.: DNRF106). We also gratefully acknowledge the support by the Spanish Ministry of Economy Competitiveness (MINECO) grant AYA-2016-76378-P. SS-D acknowledges funding by the Spanish MCIU (projects
AYA2015-68012-C2-01 and SEV-2015-0548) and the Gobierno de Canarias (project ProID2017010115)
\end{acknowledgements}

\bibliographystyle{mfa}
\bibliography{conductor.bib}

\newpage
\onecolumn
\begin{appendices}

\section{SODA database table}
\label{A1}
The database table required for the ``Conductor'' to select objects on the Central server in Aarhus (SODA) is shown below and is a combination of a number of individual tables in SODA which are combined in a table ``view''. In order to keep the database structure as general and flexible as possible, in the event of new and potentially very different instruments, all instrument-specific details are stored in the JSON structure ``constraints''. For simplicity, these are then ``unpacked'' in the table view, but it still allows us to add or change instruments without having to change the database structure.\\

\begin{table}[h!]
  \begin{center}
    \caption{SODA database table.}
    \begin{tabular}{l|c|l}
      \textbf{Table entry name} & \textbf{Input type} & \textbf{Comment}\\
      \hline
      id & Integer & Unique identifier of entry\\
      projectid & Integer & Unique project identifier\\
      starid & Integer & Unique star identifier\\
      instrument & Character & Unique instrument name\\
      schedule\_type  & Character & Schedule type name	\\
      obs\_mode  & Character & Observing mode identifier	\\
      exp\_time & Real  & Exposure time of CCD detector 	\\
      nr\_exp & Integer & Number of exposures in OR	\\
      constraints & JSON & 	Instrument and OR specifics settings\\
      readout\_mode & Smallint & Unique identifier specifying read out mode of CCD	\\
      slit & Smallint & Unique identifier of the different slits in the spectrograph	\\
      nr\_target\_exp & Smallint & Number of target spectre inside ThAr sandwich	\\
      nr\_thar\_exp & Smallint & Number of ThAr spectre before and after the target spectre	\\
      timecritical\_tstart & Timestamp  & Start time of time critical observation	\\
      timecritical\_tend & Timestamp  & Stop time of time critical observation	\\
      periodical\_deltat & Double precision & Time between observations for relevant schedule type	\\
      filler\_nr\_shots & Smallint & Number of visits for a filler observation	\\
      min\_altitude & Real & Minimum altitude of the observation - airmass	\\
      priority & Smallint  & Priority of target within the project - PI defined	\\
      projectname & Character & Unique project name  	\\
      title & Character & Project title	\\
      period & Smallint  & Unique period identifier	\\
      active & Boolean & Project state	\\
      secret & Boolean & Flag for visible state on web page 	\\
      requires\_input  & Boolean & Phase 1 flag 	\\
      pi     & Smallint & Unique identifier of project PI. 	\\
      project\_priority  & Smallint  & Project priority given by the SONG TAC	\\
      pi\_first\_name & Character & First name of project PI 	\\
      pi\_last\_name & Character & Last name of project PI 	\\
      object\_name & Character & SIMBAD resolvable object name	\\
      ra & Double precision & Right ascension of object	\\
      decl & Double precision & Declination of object	\\
      pm\_ra  & Real & Proper motion in right ascension of object 	\\
      pm\_decl  & Real & Proper motion in declination of object 	\\
      vmag & Real & Object magnitude in V-band	\\
      last\_observed & Timestamp & Last successful observation of object in project 	\\
      last\_req\_no  & Integer & Unique identifier of last OR 	\\
      last\_obs\_attempt & Timestamp & Last attempt to observe the target in project	\\
      total\_times\_observed & Integer & Total number of target observations in project	\\
      target\_active & Boolean & Object specific active state	\\
                                
    \end{tabular}
       \label{tab:table1}
  \end{center}
\end{table}

\newpage
\section{Conductor configuration}
\label{A2}
The Conductor needs to know a few site specific details to run as optimum as possible. These specifics are collected in a configuration file and a subsample of the parameters for the Tenerife site is shown in Table~\ref{tab:table2}. Details on name of log file, database table names, who to send notifying e-mails to etc. was not found relevant to present here and was therefore omitted.\\
Readout times, slewing time and overheads are mean values which needs to be set in a cleaver way for the Conductor to predict in advance when the ongoing OR will end. The time at which a new target should be inserted is roughly defined by the possible delay in the Internet traffic and a uncertainty of the calculated end time of the ongoing OR.  

\begin{table}[h!]
  \begin{center}
    \caption{Subsample on the Conductor configuration file content (Python)}
    \begin{tabular}{l|c|l}
      \textbf{Parameter} & \textbf{Value} & \textbf{Comment}\\
      \hline
      check\_time & 10.0 & Check for new target every 10 seconds\\
      project\_critical\_type & \multicolumn{2}{l}{$[$rv-standard, large\_program, periodical, filler, backup$]$ | Ordered list}\\
      insert\_time\_before & 120 & Time before detected gap starts and a new OR should be inserted.\\
      wind\_alt\_prime\_limit & 25 & Altitude limit in medium wind for large program targets\\
      wind\_alt\_other\_limit & 30 & Altitude limit for other targets - dome constraint\\
      obs\_lat & 28.2983 & Latitude of the observatory\\
      obs\_lon & -16.5094 & Longitude of the observatory\\
      obs\_elev & 2400 & Elevation of the observatory\\
      telescope\_min\_altitude & 16.0 & Minimum altitude of the telescope - telescope constraint\\
      tel\_dist\_to\_moon & 5.0 & Minimum distance to the Moon\\
      max\_alt\_auto & 82.0 & Maximum altitude for start position - pointing model constraint\\
      seeing\_limit\_s5 & 1.7 & Seeing limit for faint objects (slit 5)\\
      seeing\_limit\_s6 & 1.5 & Seeing limit for faint objects (slit 6)\\
      seeing\_limit\_s8 & 1.2 & Seeing limit for faint objects (slit 8)\\  
      vmag\_seeing\_limit & 7.5 & When to use the above seeing limits\\
      obs\_sun\_alt & -6 & Maximum altitude of Sun when observations are allowed\\   
      readoutsec & 4 & The readout time fraction in seconds of CCD\\
      readoutms & 210 & The readout time fraction in milliseconds\\
      wind\_delay & 20 & If wind is moderate and object towards wind - delay 20 minutes\\
      telescope\_slewtime & 60 & Estimate of the mean slewing time in seconds \\
      thar\_overhead & 120 & Overhead in seconds for slewing and a ThAr spectrum\\
      
    \end{tabular}
       \label{tab:table2}
         \end{center}

\end{table}

\begin{figure}
	\centering
    \includegraphics[width=\columnwidth]{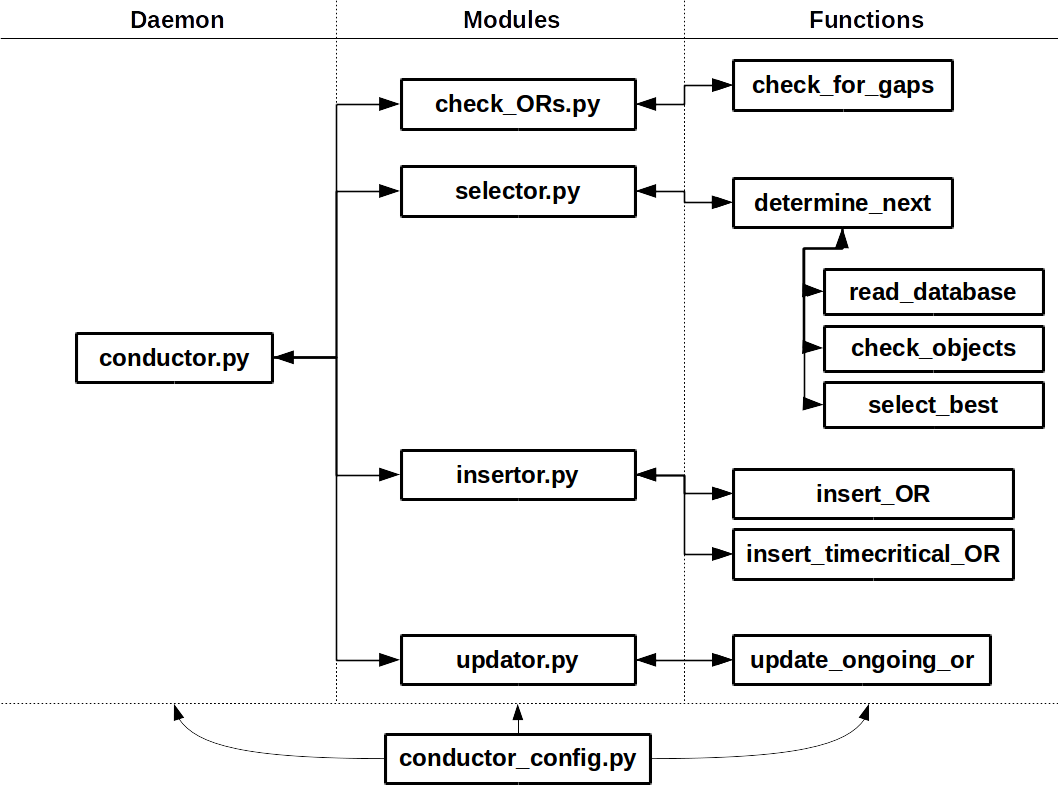}
    \caption{Schematic diagram of the Conductor software tree where only modules/functions relevant for night time observations are shown. The file ``conductor\_config.py'' will include most parameters to change for adopting to another telescope. The ``insertor.py'' needs to match the used instrument(s) and therefore might need modifications as well.}
    \label{fig:conductor_setup}
\end{figure}

\newpage

\section{OR database tables}
\label{A3}
When the Conductor has selected a target it will insert an Observing Request (OR) into a database which is replicated to the sites. Table~\ref{tab:table3} shows an example of the database table used for observations with the SONG spectrograph.\\
Together with each inserted OR a corresponding status entry is created in a database on site which will be replicated to Central. The status table is shown in Table~\ref{tab:table4}.

\begin{table}[h!]
  \begin{center}
    \caption{OR database table.}
    \begin{tabular}{l|c|l}
      \textbf{Table entry name} & \textbf{Input type} & \textbf{Comment}\\
      \hline
      req\_no & Integer & Unique identifier of OR\\
      req\_prio & Integer & Local priority of OR\\
      soda\_id & Integer & Unique identifier of SODA selection\\
      nr\_multi & Integer & Factor to allow more than 1000 spectre per OR\\
      right\_ascension & Angle360  & Right ascension of object	\\
      declination & Angle\_90to90 & Declination of object\\
      ra\_pm & Real & Proper motion in right ascension of object \\
      dec\_pm & Real & Proper motion in declination of object 	\\
      epoch & Real & Epoch of Catalogue \\
      magnitude & Real & Object magnitude in V-band	\\
      object\_name  & Character & SIMBAD resolvable object name		\\
      imagetype & Character & Type of spectra; STAR, BIAS, FLAT etc.	\\
      observer & Character & Name of the PI\\
      project\_name & Character  & Unique project name	\\
      project\_id & Integer  & Unique project identifier	\\
      exp\_time & Double precision & Exposure time of CCD detector	\\
      x\_bin & Integer & Binning of CCD in X-direction	\\
      y\_bin & Integer & Binning of CCD in Y-direction	\\
      x\_begin & Integer  & First pixel of CCD area to read out in X-direction\\
      y\_begin & Integer & First pixel of CCD area to read out in Y-direction\\
      x\_end & Integer & Last pixel of CCD area to read out in X-direction\\
      y\_end & Integer  & Last pixel of CCD area to read out in Y-direction	\\
      no\_exp & Integer & Number of spectre in OR\\
      no\_target\_exp & Integer & Number of target spectre inside ThAr sandwich\\
      no\_thar\_exp & Integer & Number of ThAr spectre before and after the target spectre	\\
      amp\_gain  & Integer & Amplifier gain of the CCD\\
      readoutmode  & Integer  & Read out mode of the CCD\\
      iodine\_cell & Integer & Position of the iodine cells\\
      obs\_mode & Character & Observing mode: ThAr, Iodine or Template\\
      slit & Integer & Unique identifier of the slit to use		\\
      start\_window  & Timestamp & Allowed start time of observing window\\
      stop\_window  & Timestamp & Stop time of the observing window\\
      acq\_spectre & Integer & Number of acquired spectre in OR\\
      on\_hold  & Real & On hold time in minutes	\\
      comment & Character & Comment from system on OR\\
      o\_star & Character & O-star to observe if template observation\\
      site & Integer & Unique site identifier	\\
      ins\_at & Timestamp & Time of OR insertion\\                               
    \end{tabular}
       \label{tab:table3}
  \end{center}
\end{table}

\begin{table}[h!]
  \begin{center}
    \caption{OR status database table.}
    \begin{tabular}{l|c|l}
      \textbf{Table entry name} & \textbf{Input type} & \textbf{Comment}\\
      \hline
      req\_no & Integer & Unique identifier of OR\\
      status & Character & Status of the OR: ``wait'', ``done'', ``abort'', ``exec'', ``unknown''\\
      ins\_at & Timestamp & Time at last status change\\
             
    \end{tabular}
       \label{tab:table4}
  \end{center}
\end{table}

\end{appendices}
\end{document}